\documentclass[12pt]{article}

\usepackage{graphicx}
\usepackage{graphics}
\usepackage{color}
\usepackage{ragged2e}
\usepackage{lipsum}
\usepackage{amsmath}

\def\beq{\begin{equation}}
\def\eeq{\end{equation}}
\def\bea{\begin{eqnarray}}
\def\eea{\end{eqnarray}}

\def\roughly#1{\mathrel{\raise.3ex\hbox
{$#1$\kern-.75em\lower1ex\hbox{$\sim$}}}}

\def\sla#1{\raise.15ex\hbox{$/$}\kern-.57em #1}

\def \cB{{\cal B}}

\def\bsmumu{b \to s \mu^+ \mu^-}
\def\bsee{b \to s e^+ e^-}

\def\bsmumu{b \to s \mu^+ \mu^-}
\def\bsee{b \to s e^+ e^-}
\def\bstautau{b \to s \tau^+ \tau^-}
\def\bsll{b \to s \ell^+ \ell^-}
\def\bsuu{b \to s {\bar u} u}
\def\bsdd{b \to s {\bar d} d}

\def\tT{{\widetilde{T}}}
\def\tC{{\widetilde{C}}}
\def\tA{{\widetilde{A}}}
\def\tP{{\widetilde{P}}}

\begin{document}

\begin{flushright}
UdeM-GPP-TH-24-301 \\
\end{flushright}

\begin{center}
\bigskip
{\Large \bf \boldmath Anomalies in Hadronic $B$ Decays\footnote{Talk given at Gravity, Strings and Fields:
    A Conference in Honour of Gordon Semenoff, July 24-28, 2023, Centre de
    recherches math\'ematiques, Montr\'eal, Canada}} \\
\bigskip
\bigskip
{\large
David London\footnote{london@lps.umontreal.ca}
}
\end{center}

\begin{center}
{\it Physique des Particules, Universit\'e de Montr\'eal,}\\
{\it 1375, ave Th\'er\`ese-Lavoie-Roux, Montr\'eal, QC, Canada H2V 0B3}
\end{center}

\begin{center}
\bigskip (\today)
\vskip0.5cm {\Large Abstract\\} \vskip3truemm
\parbox[t]{\textwidth}{In this talk, I describe a global fit to $B \to PP$ decays, where $B =
\{B^0, B^+, B_s^0\}$ and the pseudoscalar $P = \{\pi, K\}$, under the
assumption of flavour SU(3) symmetry [SU(3)$_F$]. It is found that the
individual fits to $\Delta S=0$ or $\Delta S=1$ decays are good, but
the combined fit is very poor: there is a $3.6\sigma$ disagreement
with the standard model. (This is quite a bit larger than the anomaly
in $\bsll$ transitions.) This discrepancy can be removed by adding
SU(3)$_F$-breaking effects, but 1000\% SU(3)$_F$ breaking is required,
considerably more than the $\sim 20\%$ breaking of $f_K/f_\pi - 1$.
These results are rigorous, group-theoretically -- no theoretical
assumptions have been made. But when one adds a single assumption
motivated by QCD factorization, the discrepancy grows to $4.4\sigma$.
These are the anomalies in hadronic $B$ decays. Although one cannot
yet claim that new physics is present, it is clear that something very
unexpected is going on.}

\end{center}


\thispagestyle{empty}
\newpage
\setcounter{page}{1}
\baselineskip=14pt

\section{Beyond the Standard Model}

The development of the standard model (SM) in particle physics is one
of the great triumphs in all of physics.  The SM has made a great many
predictions, almost all of which have been verified, including the
existence of the Higgs boson. There is no question that, just like
Newton's laws, the SM is correct.

However, also like Newton's laws, the SM is not complete. There are a
number of reasons that lead to this conclusion. The ``philosophical''
reasons include the large number of arbitrary parameters in the SM,
the many unanswered questions, the fact that not all forces are
unified, etc. There are also practical reasons, and these are probably
more important: the SM has no explanation for certain physics
observations such as neutrino masses, dark matter, and the
matter-antimatter asymmetry in the universe.

We therefore conclude that there must be physics beyond the SM. Most
research in particle physics today, both experimental and theoretical,
focuses on the search for this {\it new physics}.

Is this really true? Gordon Semenoff is best known for his work on
various aspects of field theory, such as theories in $2+1$ dimensions,
super-Yang-Mills theories, finite-temperature field theory, etc. One
can hardly say that the focus of this research is new physics. On the
other hand, a search through INSPIRE reveals that, since 2014, Gordon
also has nine publications with the MoEDAL experiment (in addition to
his work on field theory).  What is this? According to its CERN
website \cite{MoEDAL}, ``The MoEDAL-MAPP experiment is designed to
search for hypothetical exotic particles that would point to new
physics beyond the Standard Model of particle physics.'' So even
Gordon Semenoff is participating in the search for new physics.

\section{Searching for New Physics (Experiment)}

Experiments look for new physics (NP) in two distinct ways. In direct
searches, one aims to produce new particles in high-energy collisions,
and then to detect their presence via their decay products. If
something is seen, we will not only have found NP, we will also know
what it is. On the other hand, in indirect searches, one uses the fact
that the new particles may contribute virtually to certain observables
in some rare low-energy processes. The signal of this would then be a
discrepancy between the measured value of the observable and its SM
prediction. Here, if a disagreement is found, we can infer that NP is
present, but we will have no idea what it is.

The LHC has been performing direct searches for NP for many years.
Unfortunately, to date no new particles have been observed. This
suggests that the NP, whatever it is, must be heavy.  In this case,
the most promising way of finding NP may well be via indirect signals.

\section{Describing New Physics (Theory)}

In general, there are two ways of describing NP. Traditionally, this
was done via model building. Here a NP model is constructed by
extending the SM, either by adding new symmetries (i.e., new gauge
bosons), and/or by adding new fermions or scalars. Given a model, one
can compute the effects of these new particles on various
observables. Note that one can also consider ``simplified models,'' in
which one considers the addition of one or more NP particles, without
building a complete model.

The modern approach to describing NP uses effective field theories
(EFTs). The canonical example of an EFT is the Fermi theory. Here, the
weak interactions are described as a direct interaction among four
fermions. Later this was recognized as the low-energy limit of an
interaction mediated by the exchange of a $W$ boson. That is, the
Fermi interaction is obtained when the $W$ is ``integrated out.''

This can be generalized: the low-energy effective field theory (LEFT)
is obtained when all heavy SM particles ($W$, $Z$, $t$, $H$) are
integrated out (see, {\it e.g.}, Ref.~\cite{Jenkins:2017jig}). The
LEFT Hamiltonian obeys the unbroken $SU(3)_C \times U(1)_{em}$
symmetry, and is applicable below the weak scale.  It contains
operators of all dimensions, in which the coefficients of these
operators obey a power series in $1/M_W$.

This idea can also be applied to NP. We know the NP is heavy, so that
we can integrate it out, generating the SM effective field theory
(SMEFT) (see, {\it e.g.},
Refs.~\cite{Buchmuller:1985jz,Brivio:2017vri}). The SMEFT Hamiltonian
obeys the unbroken $SU(3)_C \times SU(2)_L \times U(1)_Y$ symmetry,
and is applicable below the NP scale [O(TeV)] and above the weak
scale. Once again, there are operators of all dimensions; their
coefficients obey a power series in $1/\Lambda_{\rm NP}$. (Most of the
dimension-4 terms correspond to the SM.)

EFTs are energy-dependent, i.e., the Hamiltonians change as one goes
from one energy scale to another. This evolution is described by the
renormalization-group equations. This means that (i) the values of the
coefficients vary with energy, and (ii) the operators mix as we change
energy scales. (This second point is very important!)

The bottom line is that all this must be taken into account when
looking for a (high-energy) NP explanation of a low-energy discrepancy
with the SM.

\section{\boldmath Semileptonic $\bsll$ Anomalies}

For over ten years, the values of a number of observables in processes
involving $\bsmumu$ transitions have shown significant disagreements
with the predictions of the SM. Examples of such processes include $B \to
K^{(*)} \mu^+ \mu^-$, $B_s^0 \to \phi \mu^+ \mu^-$, etc. In addition,
the ratios
\beq
R_{K^{(*)}} \equiv \frac
{\cB(\bar{B} \to K^{(*)} \mu^+ \mu^-)}{\cB(\bar{B} \to K^{(*)} e^+ e^-)}
\eeq
were measured, and for a number of years, their values differed
significantly from 1 (the SM prediction). The most compelling
explanation was that NP affects $\bsmumu$, but not $\bsee$
\cite{London:2021lfn}.

In December, 2022, LHCb announced that they now find $R_{K^{(*)}}
\simeq 1$, in agreement with the SM
\cite{LHCb:2022qnv,LHCb:2022vje}. This implies that the NP contributes
equally to $\bsmumu$ and $\bsee$
\cite{Greljo:2022jac,Alguero:2023jeh,Hurth:2023jwr}; this is referred
to as ``lepton-flavour-universal NP''.

It is possible to construct models to explain this. For example, a
$Z'$ boson could contribute to $\bsll$ at tree level. The $Z'$ must
have an FCNC coupling to ${\bar s} b$ and must couple equally to $e^+
e^-$ and $\mu^+ \mu^-$.

On the other hand, it is more informative to examine the problem from
a model-independent (EFT) point of view.  Here's what we know.  The
$\bsll$ anomalies are observed at the scale $m_b$. The fits
\cite{Greljo:2022jac,Alguero:2023jeh,Hurth:2023jwr} show that the
preferred solution requires equal new contributions to the LEFT
operators ${\bar s} \gamma^\mu P_L b {\bar\mu} \gamma_\mu \mu$ and
${\bar s} \gamma^\mu P_L b {\bar e} \gamma_\mu e$. We therefore have
the following question. These LEFT operators are independent; their
coefficients are a-priori unrelated.  Can we find SMEFT operators that
generate equal coefficients for these operators when the theory is run
down from $\Lambda_{\rm NP}$ to $m_b$ using the renormalization group?

\begin{figure}[!htbp]
\begin{center}
\includegraphics[width=0.35\textwidth]{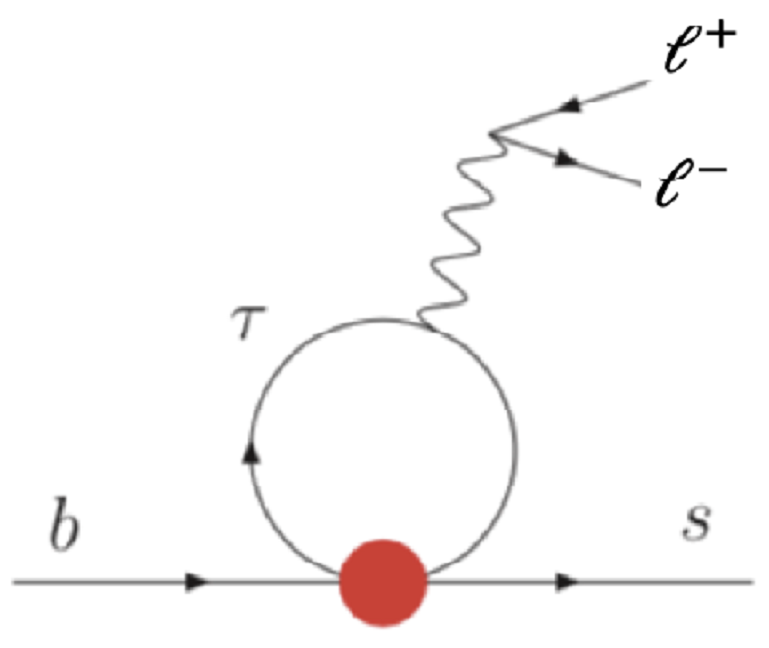}
\end{center}
\caption{Generation of the operators ${\bar s} \gamma^\mu P_L b
  {\bar\mu} \gamma_\mu \mu$ and ${\bar s} \gamma^\mu P_L b {\bar e}
  \gamma_\mu e$ with equal coefficients.}
\label{LFUgen}
\end{figure}

Yes! Here is one example: Suppose there is a $\bstautau$ SMEFT
operator at scale $\Lambda_{\rm NP}$. When we run the theory down to
$m_b$, the LEFT operators ${\bar s} \gamma^\mu P_L b {\bar\mu}
\gamma_\mu \mu$ and ${\bar s} \gamma^\mu P_L b {\bar e} \gamma_\mu e$
are generated at one loop through $\gamma$ exchange, see
Fig.~\ref{LFUgen}. And since $Q_{em}^e = Q_{em}^\mu$, their
coefficients are equal!

I note in passing that there are other SMEFT operators that generate
LEFT ${\bar s} \gamma^\mu P_L b {\bar\ell} \gamma_\mu \ell$ operators
in this way; this type of operator mixing is very common. Also, SMEFT
doesn't say how these operators are produced -- that's for the model
builders to do. For example, the $\bstautau$ SMEFT operator could be
produced via a $Z'$ or a leptoquark (LQ) exchange. But models that
contain a $Z'$ or a LQ may also have other new particles. And SMEFT
operators other than $\bstautau$ may be generated when all NP
particles are integrated out, in which case the fit must be repeated.
This shows how the model-building and EFT approaches are
complementary: both must be considered when searching for a NP
explanation of a low-energy anomaly.

\section{\boldmath Hadronic $B$ Anomalies}

Returning to Fig.~\ref{LFUgen}, note that the $\gamma$ can decay to
particles other than just $e^+e^-$ and $\mu^+\mu^-$. In particular, it
can decay to ${\bar u}u$ and ${\bar d}d$. That is, on very general
grounds, one also expects NP contributions to $\bsuu$ and $\bsdd$. In
other words, we can expect to see NP effects in $\Delta S=1$ hadronic
$B$ decays!

In fact, there are several $\Delta S=1$ hadronic $B$ decays that
exhibit deviations from the SM predictions, but they have not
attracted nearly as much attention as the $\bsll$ anomalies. They
include:
\begin{enumerate}

\item The $B \to\pi K$ puzzle has been around for about 20 years
  \cite{Beaudry:2017gtw}. The amplitudes for the four decays
  $B^+\to\pi^0 K^+$, $B^+\to\pi^+ K^0$, $B^0\to\pi^- K^+$, and
  $B^0\to\pi^0 K^0$ form an isospin quadrilateral. Nine observables
  have been measured in these decays, and a combined analysis shows
  that these results are not entirely consistent with one
  another. There is a $\sim 2\sigma$ discrepancy with the SM.

\item The U-spin puzzle \cite{Bhattacharya:2022akr}: U-spin symmetry
  ($d \leftrightarrow s$) relates the six decays $B_{d,s}^0 \to P^\pm
  P^{\prime\mp}$, where $P$ and $P'$ are each $\pi$ or $K$. 12
  observables have been measured in these decays; once again, a
  combined analysis reveals a certain inconsistency. One can explain
  the data by allowing for U-spin-breaking effects, but we find that
  100\% U-spin breaking is required, much larger than the $\sim
  25$-30\% expected in SM. This results hints at NP in $B_s^0 \to K^+
  K^-$.

\item Three puzzles involving $B_s^0 \to K^0 {\bar K}^0$
  \cite{Amhis:2022hpm}: the decay $B_s^0 \to K^0 {\bar K}^0$ is
  related to $B_s^0 \to K^+ K^-$ by isospin, to $B_d^0 \to K^0 {\bar
    K}^0$ by U spin, and to $B^+ \to \pi^+ K^0$ by virtue of having
  the same quark-level decay, ${\bar b} \to {\bar s} d {\bar d}$).  In
  all three cases, there is a disagreement of $\sim 2$-$3\sigma$ in
  the branching ratios of the pair of decays.

\end{enumerate}
The point here is that we have five different sets of $B$ decays that
exhibit a significant discrepancy with the SM. All are $\Delta S = 1$
decays, i.e., they involve the $b \to s$ transition, just like the
semileptonic $\bsll$ anomalies. This is very intriguing.

This talk was given on July 27, 2023. At the time, I noted that
members of my group were examining these hadronic $B$ anomalies, and I
presented some preliminary results. Since then, the analysis has been
completed and the paper written up in Ref.~\cite{Berthiaume:2023kmp}.
Below I summarize the analysis and its results.

All of the above processes are $B\to PP$ decays, where $B = \{B^0,
B^+, B_s^0\}$, and the pseudoscalar $P = \{ \pi, K \}$. These decays
are all related to one another by flavour SU(3) symmetry [SU(3)$_F$].
One can therefore express all amplitudes in terms of SU(3)$_F$ reduced
matrix elements (RMEs). By performing a global fit to all the $B \to
PP$ observables under the assumption of SU(3)$_F$, these puzzles can
be combined, and one can quantify just how well (or poorly) the SM
explains the data.

It is straightforward to show that there are a total of seven
independent RMEs. Defining $\lambda_{q'}^{(q)} \equiv V_{q'b}^*
V_{q'q}$, $q=d,s$, $q' = u,c,t$, two RMEs are proportional to
$\lambda_{u}^{(q)}$, two are proportional to $\lambda_{t}^{(q)}$, and
three RMEs involve both $\lambda_{u}^{(q)}$ and $\lambda_{t}^{(q)}$.
If SU(3)$_F$ is unbroken, these RMEs are the same for $\Delta S=0$
($q=d$) and $\Delta S=1$ ($q=s$) decays. However, they can be
different if SU(3)$_F$-breaking effects are allowed.

Many years ago, it was shown that an equivalent description of these
SU(3)$_F$ amplitudes is provided by topological amplitudes (quark
diagrams) \cite{Gronau:1994rj,Gronau:1995hn}. There are eight
topologies.  The tree ($T$), color-suppressed tree ($C$), annihilation
($A$), and $W$-exchange ($E$) diagrams are associated with
$\lambda_u^{(q)}$. The electroweak penguin ($P_{EW}$) and
color-suppressed electroweak penguin ($P^C_{EW}$) are associated with
$\lambda_t^{(q)}$. The penguin ($P$) comes in two types -- $P_{uc}$
and $P_{tc}$ are associated with $\lambda_{u}^{(q)}$ and
$\lambda_{t}^{(q)}$, respectively -- and similarly for the
penguin-annihilation ($PA$) diagram. These ten diagrams appear in
seven linear combinations in the amplitudes; these are related to the
seven RMEs.

One advantage of diagrams over RMEs is that it is easy to write
amplitudes in terms of diagrams -- no (tricky) SU(3)$_F$
Clebsch-Gordan coefficients need to be computed. Another advantage is
that one can estimate the relative sizes of different diagrams. For
example, it has been argued that $E$, $A$ and $PA$ can be neglected
compared to the other diagrams because they involve an interaction
with the spectator quark \cite{Gronau:1994rj,Gronau:1995hn}. But this
is also problematic: results that use dynamical assumptions such as
this are not rigorous group-theoretically. In our analysis, we use
diagrams, but we make no assumptions about their sizes. These are
fixed only by the data.

There are eight $\Delta S = 0$ $B \to PP$ decays. These are
parametrized by 7 (effective) diagrams, corresponding to 13 unknown
theoretical parameters (7 magnitudes and 6 relative strong
phases). But 15 observables have been measured. These include
CP-averaged branching ratios, direct CP asymmetries, and indirect CP
asymmetries.  With more observables than unknowns, a fit can be done
with only $\Delta S = 0$ decays. We find $\chi_{\rm min}^2/{\rm
  d.o.f.} = 0.35/2$, for a $p$-value of 0.84. This is an excellent
fit.

There are also eight $\Delta S = 1$ $B \to PP$ decays, again
parametrized by 7 (effective) diagrams. For these decays, there are
again 15 observables, so a fit can be performed. We find $\chi_{\rm
  min}^2/{\rm d.o.f.} = 1.8/2$, for a $p$-value of 0.4. This is a very
good fit.

But here's the kicker. Under SU(3)$_F$, $\Delta S = 0$ and $\Delta S =
1$ decays are the same. We can then perform a combined fit, including
all the data. But when such a fit is performed, we find that
$\chi_{\rm min}^2/{\rm d.o.f.} = 43.8/17$, for a $p$-value of $3.6
\times 10^{-4}$. This corresponds to a discrepancy with the SM at the
level of $3.6\sigma$.

Once again, I stress that no dynamical assumptions have been made
regarding the diagrams. This result is completely rigorous from a
group-theoretical point of view.

Note that the $3.6\sigma$ is not the ``pull,'' which has become
popular in the analysis of the $\bsll$ anomalies. The pull, which
often reaches 4-5$\sigma$ or even larger, indicates how much better
than the SM a particular NP model describes the data. But it says
nothing about how well the SM explains the data. In fact, even when
the measured values of $R_{K^{(*)}}$ differed from 1, the overall
discrepancy with the SM was only around $2.3\sigma$. That is, even
though the analyses of the $\bsmumu$ anomalies found NP scenarios with
pulls of $\sim 6\sigma$, in actual fact, the SM did not do such a poor
job of explaining the data.  So the hadronic $B$ anomalies are quite a
bit more problematic.

Of course, the analysis has assumed unbroken SU(3)$_F$ symmetry.
However, we know that SU(3)$_F$ is, in fact, broken. For example,
$f_K/f_\pi - 1 = ~\sim 20\%$; this is typically taken to be the
expected size of SU(3)$_F$ breaking in the SM. Could the addition of
such SU(3)$_F$-breaking effects remove the discrepancy?

Fortunately, the fit contains enough information to address this
question. Above, we found an excellent fit when one considers only
$\Delta S = 0$ decays and a very-good fit for $\Delta S = 1$ decays.
In both cases, the best-fit values of the (effective) diagrams can be
extracted. These are shown in Table \ref{tab:params_diags}.  Denoting
the $\Delta S=0$ and $\Delta S=1$ diagrams by $D$ and $D'$,
respectively, in the SU(3)$_F$ limit, $D_i = D'_i$. However, we see in
the Table that one typically finds $|D_i'/D_i| = O(10)$, which
corresponds to 1000\% SU(3)$_F$ breaking! That is, if the SM really
does explain the data, then the breaking of the flavour SU(3)$_F$
symmetry must be at an unexpectedly large level.

\begin{table}[!htbp]
\begin{center}
\begin{tabular}{|c|c|c|c|c|} \hline
& $|\tT|$ & $|\tC|$ & $|\tP_{uc}|$ & $|\tA|$\\ \cline{2-5}
Fit & $4.0\pm0.5$ & $6.6\pm0.7$ & $3\pm4$ & $6\pm5$ \\ \cline{2-5}
$\Delta S = 0$ & $|{\widetilde{PA}}_{uc}|$ & $|P_{tc}|$ & $|PA_{tc}|$ & \\ \cline{2-5}
& $0.7\pm0.8$ & $0.8\pm0.4$ & $0.2\pm0.4$ & \\ \hline\hline         
& $|\tT'|$ & $|\tC'|$ & $|\tP'_{uc}|$ & $|\tA'|$ \\ \cline{2-5} 
Fit & $48\pm14$ & $41\pm14$ & $48\pm15$ & $81\pm28$ \\ \cline{2-5}
$\Delta S = 1$ & $|{\widetilde{PA}}'_{uc}|$ & $|P'_{tc}|$ & $|PA'_{tc}|$ & \\ \cline{2-5}
& $7\pm4$ & $0.78\pm0.16$ & $0.24\pm0.04$ & \\ \hline\hline
& $|\tT'/\tT|$ & $|\tC'/\tC|$ & $|\tP'_{uc}/\tP_{uc}|$ & $|\tA'/\tA|$ \\ \cline{2-5}
& $12\pm4$ & $6.6\pm2.2$ & $16\pm22$& $14\pm13$ \\ \cline{2-5}
& $|{\widetilde{PA}}'_{uc}/{\widetilde{PA}}_{uc}|$ & $|P'_{tc}/P_{tc}|$ & $|PA'_{tc}/PA_{tc}|$ & \\ \cline{2-5}
& $10\pm13$ & $0.97\pm0.52$& $1.3\pm2.7$ & \\ \hline
\end{tabular}
\end{center}
    \caption{Best-fit values of the magnitudes of the effective
      diagrams for the $\Delta S=0$ and $\Delta S=1$ fits, as well as
      their ratios. For definitions of the diagrams, see
      Ref.~\cite{Berthiaume:2023kmp}.
    \label{tab:params_diags}}
\end{table}

But there's more. Until now, no theoretical assumptions about the
diagrams have been made; the results are rigourous, from a
group-theoretical point of view. Now, two of the diagrams are $T$ and
$C$, respectively the colour-allowed and colour-suppressed tree
diagrams. Naively we expect $|C/T| = 1/3$, simply by counting
colors. In QCD factorization, this ratio is computed for $B \to \pi K$
decays ($\Delta S=1$), and $|C/T| \simeq 0.2$ is found
\cite{Beneke:2001ev,Bell:2007tv,Bell:2009nk,Beneke:2009ek,Bell:2015koa}. To
a good approximation, $|\tC/\tT| \simeq |C/T|$. If we fix
$|\tC^{(\prime)}/\tT^{(\prime)}|$ to 0.2 and redo the fits, we now
find that, when the $\Delta S=0$ and $\Delta S=1$ decays are combined
(i.e., perfect SU(3)$_F$ symmetry is assumed), the best fit has
$\chi_{\rm min}^2/{\rm d.o.f.} = 55.8/18$, for a $p$-value of $9.4
\times 10^{-6}$. The discrepancy with the SM has grown significantly:
it is now $4.4\sigma$!

These are the anomalies in hadronic $B$ decays. It is far too early to
claim that we are seeing signs of new physics, but it is clear that
something very unexpected is going on. This needs to be investigated
further.

\bigskip
\noindent
{\bf Acknowledgments}: This work was financially supported in part by
NSERC of Canada.

%

\end{document}